\documentclass[12pt]{article}
\usepackage{graphicx,citesort,amsmath,amssymb}
\def \app{D_{\pi \pi}}

\def \b{{\cal B}}
\def \bb{\bar{\cal B}}
\def \bea{\begin{eqnarray}}
\def \beq{\begin{equation}}
\def \bl{\bar \lambda}

\def \cn{Collaboration}

\def \eea{\end{eqnarray}}
\def \eeq{\end{equation}}
\def \ite{{\it et al.}}

\def \s{\sqrt{2}}

\def \sx{\sqrt{6}}
\renewcommand{\thetable}{\Roman{table}}
\renewcommand{\thesection}{\Roman{section}}
\begin{document}
\begin{flushright}
SLAC-PUB-12043\\
EFI 06-17 \\
hep-ph/0608085 \\
August 2006 \\
\end{flushright}
\renewcommand{\thesection}{\Roman{section}}
\renewcommand{\thetable}{\Roman{table}}
\centerline{\bf UPDATED BOUNDS ON CP ASYMMETRIES IN $B^0 \to \eta'
K_{S}$ AND $B^0 \to \pi^0 K_{S}$
\footnote{Research supported in
part by the US Department of Energy, contract DE-AC02-76SF00515.}}
\medskip
\vskip3mm \centerline{Michael Gronau$^{a,b}$, Jonathan L.
Rosner$^c$, and Jure Zupan$^{d}$}
\medskip
\vskip3mm
\centerline{$^a$\it Physics Department, Technion -- Israel Institute of
Technology}
\centerline{\it 32000 Haifa, Israel}
\medskip
\centerline{$^b$\it Stanford Linear Accelerator Center, Stanford
University} \centerline{\it Stanford, CA 94309, USA}
\medskip
\centerline{$^c$\it Enrico Fermi Institute and Department of
Physics, University of Chicago} \centerline{\it Chicago, Illinois
60637}
\medskip
\centerline{$^d$\it Carnegie-Mellon University, Pittsburgh, PA
15213, and} \centerline{\it J.~Stefan Institute, Jamova 39, P.O. Box
3000} \centerline{\it 1001 Ljubljana, Slovenia}
\bigskip
\begin{quote}
New rate measurements of $B^0$ decays into $\pi^0\pi^0,~\pi^0\eta,~\pi^0\eta',
~\eta\eta,~\eta\eta',~\eta'\eta'$ and $K^+ K^-$ are used in conjunction with
flavor SU(3) to constrain the coefficients $S$ and $C$ of $\sin\Delta mt$ and 
$\cos\Delta mt$ in the time-dependent CP asymmetries of $B^0 \to \eta' K_S$
and $B^0 \to \pi^0 K_S$.  Experimental values of $S_{\eta' K}$ are now seen to
be closer to the Standard Model expectations, fully consistent with the new
improved bounds.
\end{quote}
\leftline{\qquad PACS codes:  12.15.Hh, 12.15.Ji, 13.25.Hw, 14.40.Nd}
\bigskip

\section{INTRODUCTION}

Time-dependent CP asymmetries in $B^0$ decays to CP eigenstates dominated by
the $b \to s$ penguin amplitude have for several years been fertile ground for
exploring signatures of new physics \cite{NP}.  The decay $B^0 \to \eta' K_S$,
as one example, attracted attention because of the possible deviation of the
coefficient $S_{\eta' K_S}$ of the $\sin \Delta mt$ term from its predicted
value of $\sin 2 \phi_1 = \sin 2 \beta$, where $\beta \equiv \arg(-V_{tb}
V^*_{td}V_{cd}V_{cb}^*)$ is one of the phases in the standard unitary triangle
constructed from the Cabibbo-Kobayashi-Maskawa (CKM) matrix elements.  A value
$\sin 2 \beta = 0.674 \pm 0.026$ is extracted from $B \to J/\psi K_{S,L}$
decays \cite{Hazumi},

In Refs.\ \cite{Gronau:2004hp} and \cite{Gronau:2003kx} correlated bounds on
$S$ and $C$ parameters in time-dependent decays $B^0 \to \eta' K_S$ and $B^0
\to \pi^0 K_S$ were obtained using branching ratio measurements of
SU(3)-related $B^0$ decays.  The BaBar Collaboration now has updated its
measurements of a number of branching ratios which contributed to the bounds in
Ref.\ \cite{Gronau:2004hp}, leading to a further strengthening of these bounds
within the Standard Model.  These new results include branching ratios for $B^0
\to \eta \eta',~\eta \pi^0,~\eta' \pi^0$ based on 232 million $\bar B B$ pairs
\cite{Aubert:2006qd}, for $B^0 \to \eta \eta,~\eta' \eta'$ based on 324 million
$\bar B B$ pairs \cite{Baetaeta}, and $B^0 \to \pi^0 \pi^0$ based on 347
million $\bar B B$ pairs \cite{Bapzpz}.  Belle has also updated its branching
ratio for $B^0 \to \pi^0 \pi^0$ based on 532 million $\bar B B$ pairs
\cite{Unno}.  At the same time BaBar has presented new values for $S_{\eta' K}$
and $C_{\eta' K}$ which are closer to the predictions of the Standard Model
\cite{Aubert:2006wv}, while Belle has updated its values based on more data
\cite{Hazumi,Hara,Abe:2006nk}.  Finally, new measurements of $S_{\pi^0 K_S}$
and $C_{\pi^0 K_S}$ were presented by both Belle \cite{Hara} and BaBar
\cite{Aubert:2006ad}.  The purpose of this work is to compare the new
predictions with the new measurements.  The considerable improvements in bounds
in comparison with our earlier treatments \cite{Gronau:2004hp,Gronau:2003kx}
deserve to be noted despite the fact that we break no new theoretical
ground here.  Where otherwise unspecified we use values of branching ratios
quoted by the Heavy Flavor Averaging Group~\cite{HFAG}.

Alternative approaches for studying the asymmetries $S$ and $C$ in $b\to s$ 
penguin-dominated $B^0$ decays have been adopted in other works by calculating 
hadronic amplitudes for these processes within the frameworks of QCD 
Factorization~\cite{Beneke:2003zv}, Soft Collinear Effective Theory 
(SCET)~\cite{Williamson:2006hb} and a model for final state 
interactions~\cite{Cheng:2005bg}.

In Section II we briefly sketch the formalism for the case of time-dependent
asymmetries in $B^0 \to \eta' K_S$, referring to \cite{Gronau:2004hp} for
details, and present the new bounds on $S_{\eta' K_S}$ and $C_{\eta' K_S}$.
In Section III the updated bounds for $B^0 \to \pi^0 K_S$ will then be given.
We summarize in Section IV, stressing the fact that our bounds may
be approaching their optimum limits.

\section{BOUNDS FOR $S_{\eta' K_S}$ AND $C_{\eta' K_S}$}

For $\eta' K_S$ the asymmetry has the form \cite{MG}:
\beq
A(t) \equiv \frac{\Gamma(\bar B^0(t) \to \eta' K_S) - \Gamma(B^0(t) \to \eta'
K_S)}{\Gamma(\bar B^0(t) \to \eta' K_S) + \Gamma(B^0(t) \to \eta' K_S)}
= -C_{\eta' K} \cos(\Delta mt) + S_{\eta' K}\sin(\Delta mt)~,
\eeq
with
\beq
S_{\eta' K} \equiv \frac{2{\rm Im}(\lambda_{\eta' K})}{1 + |\lambda_{\eta'
K}|^2}~, ~~~C_{\eta' K} \equiv \frac{1 - |\lambda_{\eta' K}|^2}{1 +
|\lambda_{\eta' K}|^2}~, ~~~\lambda_{\eta' K} \equiv -e^{-2i\beta}\frac{A(\bar
B^0 \to \eta' \bar K^0)}{A(B^0 \to \eta' K^0)}~.
\eeq
We decompose the $B^0 \to \eta' K^0$ amplitude into two terms $A'_P$ and $A'_C$
containing respectively the CKM factors $V^*_{cb}V_{cs}$ and $V^*_{ub}V_{us}$
\footnote{The normalization of $A'_{P,C}$ differs from the one in
\cite{Gronau:2004hp} by $\sx$.  This normalization cancels in the results for
$S_{\eta' K}, C_{\eta'K}$.}
\beq\label{Amp}
A(B^0\to \eta' K^0) = A'_P + A'_C = |A'_P|e^{i\delta} + |A'_C|e^{i\gamma}~,
\eeq
where $\delta$ and $\gamma$ in the last equality are respectively the strong
and the weak phase. In the diagrammatic language $A'_P$ is the dominant $b \to
s$ penguin amplitude and  $A'_C$ is a color-suppressed amplitude.

The asymmetries $S_{\eta' K}$ and $C_{\eta' K}$ are
\bea
S_{\eta' K} & = & {\sin 2\beta + 2|A'_C/A'_P| \cos \delta \sin(2 \beta +
\gamma) - |A'_C/A'_P|^2 \sin(2 \alpha) \over R_{\eta'K}}~ \label{eqn:S},\\[3pt]
C_{\eta' K} & = & {2|A'_C/A'_P| \sin \delta \sin \gamma \over R_{\eta'K}}~
\label{eqn:C},\\[3pt]
R_{\eta'K} & \equiv & 1 + 2|A'_C/A'_P| \cos \delta \cos \gamma +|A'_C/A'_P|^2~.
\label{eqn:R}
\eea
The amplitudes $A'_P$ and $A'_C$ are expected to obey $|A'_C| \ll |A'_P|$
\cite{GHLR}.  If $A'_C$ were neglected one would have $S_{\pi K} = \sin 2\beta,
C_{\pi K}=0$. Keeping only linear terms in $|A'_C/A'_P|$ \cite{MG}
one would have an allowed region in the $(S_{\eta'K}, C_{\eta'K})$ plane
lying inside an ellipse centered at $(\sin 2 \beta, 0)$.  We
use the exact expressions (\ref{eqn:S})--(\ref{eqn:R}).  Bounds on $\gamma$
from global CKM analyses \cite{CKMfitter} lead to asymmetries in the
approximately elliptical regions surrounding the Standard Model point.

Using the flavor-SU(3) decomposition of Refs.\ \cite{GHLR,DZ,Chau,DGR,%
SW,Desh,GL} one can express the ratio $A'_C/A'_P$ in terms of SU(3)-related
amplitudes $A_C/A_P$ for $\Delta S = 0$ $B^0$ decays as pointed out in
\cite{GLNQ}. The bounds on $\Delta S_{\eta' K}\equiv S_{\eta' K}-\sin 2\beta$
and $C_{\eta' K}$ then arise because $A'_C=\bl A_C$ is CKM-suppressed, while
$A'_P=-\bl^{-1} A_P$ is CKM-enhanced compared to the $\Delta S=0$ amplitudes
(here $\bl=-V_{cd}/V_{cs}=0.230$).  Writing $A_{P, C}$ in terms of the $\Delta
S=0$ $B\to f$ amplitudes $A_f$
\beq\label{psc}
\Sigma_f a_fA(f) = A_P + A_C~,
\eeq
one then obtains the bounds (see \cite{Gronau:2004hp} for details)
\beq\label{boundsA'c/A'p}
\frac{|{\cal R} - \bl^2|}{1 + {\cal R}}
\le |A'_C/A'_P| \le \frac{{\cal R} + \bl^2}{1 - {\cal R}}~.
\eeq
The ratio ${\cal R}$ is
\beq\label{R}
{\cal R}^2 \equiv
\frac{\bl^2[|\Sigma_f a_fA(f)|^2 + |\Sigma_f a_f \bar A(f)|^2]}
{|A(B^0\to \eta' K^0)|^2 + |A(\bar B^0 \to \eta' \bar K^0)|^2},
\eeq
and is bounded by
\beq\label{boundR}
{\cal R} \le \bl \Sigma_f |a_f| \sqrt{\frac{\bar {\cal B}_f}{\bar {\cal B}
(\eta'K^0)}}~.
\eeq
For a given set of coefficients $a_f$, nonzero branching ratio measurements and
upper limits on CP averaged branching ratios $\bar {\cal B}_f$ provide an upper
bound on ${\cal R}$, for which the right-hand-side of (\ref{boundsA'c/A'p})
gives an upper bound on $|A'_C/A'_P|$.

\begin{table}
\caption{Branching ratios in $10^{-6}$ and $90\%$ C.L. upper limits on
branching ratios}
\begin{center}
\begin{tabular}{c c c c c c c c} \hline \hline
Mode  & $\eta' K^0$  & $\pi^0\pi^0$ & $\pi^0\eta$ &
 $\pi^0\eta'$ & $\eta\eta$ & $\eta'\eta'$ & $\eta\eta'$ \\ \hline
This work & $64.9 \pm 3.5$ & $1.31 \pm 0.21$ & $< 1.3$ &
 $1.5^{+0.7}_{-0.6}$ & $< 1.8$ & $< 2.4$ & $< 1.7$ \\
Ref.\ \cite{Gronau:2004hp} & $65.2 ^{+6.0}_{-5.9}$ & $1.9 \pm 0.5$ &
 $< 2.5$ & $< 3.7$ & $< 2.8$ & $ < 10$ & $ < 4.6$ \\ \hline \hline
\end{tabular}
\end{center}
\end{table}

Since there are more physical amplitudes $A(f)$ than SU(3) contributions, one
may form a variety of combinations satisfying (\ref{psc}).
We consider two of the cases noted in Ref.\ \cite{Gronau:2004hp}:

\begin{enumerate}
\item A combination involving pairs including $\pi^0,~\eta$ and $\eta'$ in
the final state was proposed in~\cite{GLNQ} by using a complete SU(3)
analysis, and in~\cite{CGR} by applying U-spin symmetry arguments:
\bea\label{six}
\Sigma_f a_fA(f)& = &
\frac{1}{4\sqrt{3}}A(\pi^0\pi^0) - \frac{1}{3}A(\pi^0\eta) +
\frac{5}{6\sqrt{2}}A(\pi^0\eta')
\nonumber\\
& + & \frac{2}{3\sqrt{3}}A(\eta\eta)  - \frac{11}{12\sqrt{3}}A(\eta'\eta') -
\frac{5}{3\sqrt{3}}A(\eta\eta')~.
\eea
\item Another superposition, satisfying (\ref{psc}) in the limit in which
small amplitudes involving the spectator quark may be neglected,
involves only three strangeness-conserving amplitudes:
\beq\label{three}
\Sigma_f a_fA(f) = -\frac{5}{6}A(\pi^0\eta) + \frac{1}{3\sqrt{2}}A(\pi^0\eta')
- \frac{\sqrt{3}}{2}A(\eta\eta')~.
\eeq
\end{enumerate}
The coefficients $a_f$ in these cases can be read off Eqs.~(\ref{six})
and (\ref{three}).

As mentioned before, the upper bounds for a number of the relevant decays have
been strengthened recently.  In units of $10^{-6}$, we use the value
\cite{HFAG} $\b(\eta' K^0) = 64.9$ (we ignore the error $\pm 3.5$ as in Ref.\
\cite{Gronau:2004hp}) and the 90\% c.l.\  upper limits
$\b(\pi^0 \pi^0) < 1.58$,
$\b(\pi^0 \eta) < 1.3$, $\b(\pi^0 \eta')
< 2.4$, $\b(\eta \eta) < 1.8$, $\b(\eta' \eta') < 2.4$, and $\b(\eta \eta') <
1.7$.  These inputs are compared with those used in Ref.\ \cite{Gronau:2004hp}
in Table I.  The bounds on ${\cal R}$ obtained in the above two cases are then
as follows:

\begin{enumerate}

\item Assuming exact SU(3) and applying (\ref{six}) we find, using the central
value for $\bar {\cal B}(\eta' K^0)$,
\beq\label{boundR1}
{\cal R} < 0.116~({\rm formerly}~0.18)~.
\eeq

\item Using (\ref{three}), which contains three processes, one finds
\beq\label{boundR3}
{\cal R} < 0.070~({\rm formerly}~0.10)~~.
\eeq

\end{enumerate}
The approximation involved in deriving (\ref{boundR3}),
where SU(3) breaking and small amplitudes were neglected,  is comparable
to that associated with (\ref{boundR1}) which only neglects SU(3) breaking
effects.

\begin{figure}[t]
\includegraphics[height=4.9in]{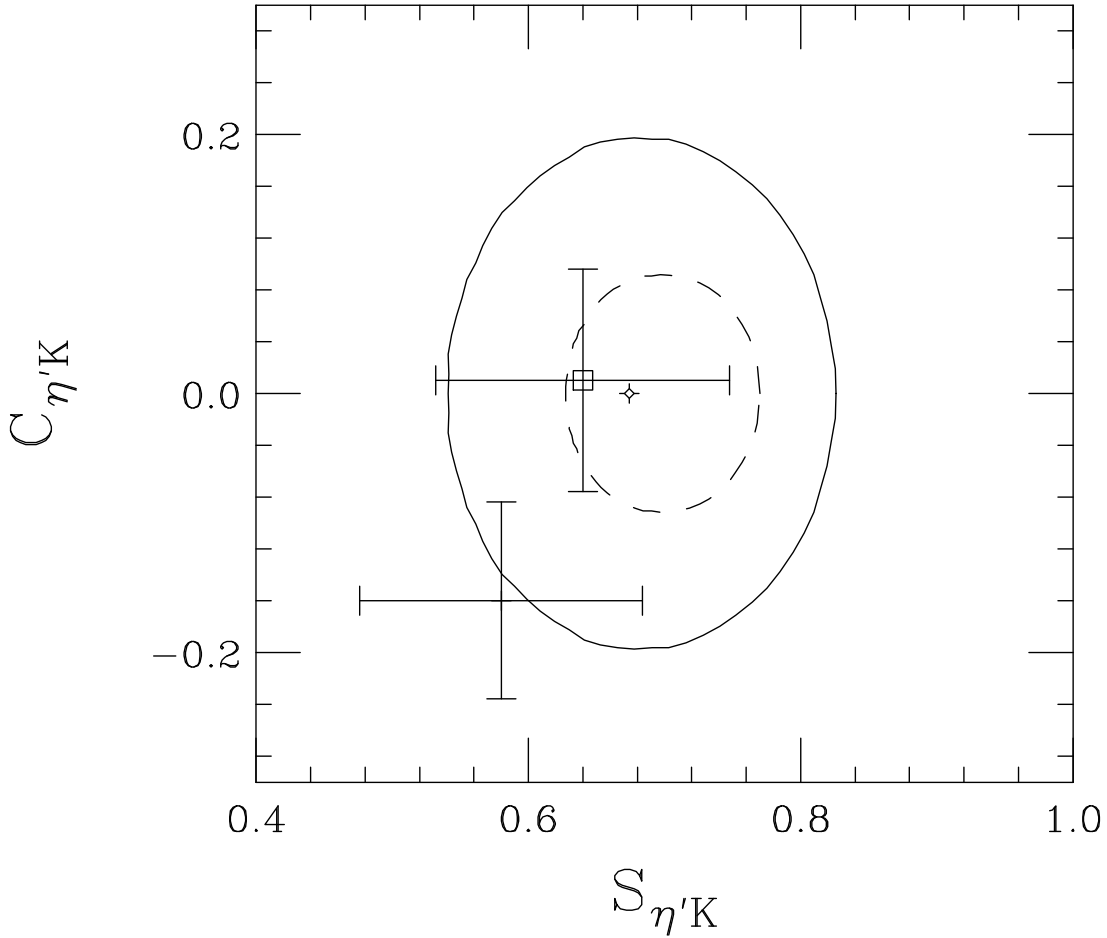}
\caption{Regions in the ($S_{\eta' K},~C_{\eta' K}$) plane satisfying
limits on the ratio $|A'_C/A'_P|$ and bounds (\ref{boundR1}) (region
enclosed by the solid curve) or (\ref{boundR3}) (region enclosed by the dashed
curve).  The small plotted point denotes $(S_{\eta' K},~C_{\eta' K}) = (\sin
2 \beta,~0)$.  The points with experimental errors denote values from BaBar
\cite{Aubert:2006wv} (plain point) and Belle \cite{Abe:2006nk} (small square).
\label{fig:maxup0608}}
\end{figure}

In order to study constraints in the ($S_{\eta' K}, C_{\eta' K}$) plane, we now
apply the upper bounds (\ref{boundR1}) and (\ref{boundR3}).  The exact
expressions (\ref{eqn:S})--(\ref{eqn:R}) imply correlated bounds on these two
quantities associated with fixed values of ${\cal R}$. We scan over $-\pi \le
\delta \le \pi$, taking a central value $\beta = 21.2^{\circ}$, values of
$\gamma$ satisfying $52^\circ \le \gamma \le 74^\circ$ \cite{CKMfitter}, and
values of $|A'_C/A'_P|$ in the range (\ref{boundsA'c/A'p}), where ${\cal R}$
satisfies the bound (\ref{boundR1}) or (\ref{boundR3}).  The bounds on
($S_{\eta' K}, C_{\eta' K}$) are shown in Fig.~1.  The small plotted point
corresponds to $(S_{\eta'K}, C_{\eta'K}) = (\sin 2\beta, 0)$ (see below). The
large plotted points correspond to the most recent results reported by BaBar
\cite{Aubert:2006wv} and Belle \cite{Abe:2006nk}.  These results are noted in
Table II.

\begin{table}
\caption{Time-dependent asymmetries in $B^0 \to \eta' K_S$.
\label{tab:etapasyms}}
\begin{center}
\begin{tabular}{c c c} \hline \hline
Parameter       & BaBar \cite{Aubert:2006wv} & Belle \cite{Abe:2006nk} \\
\hline
$S_{\eta' K_S}$ & $0.58 \pm 0.10 \pm 0.03$ & $0.64 \pm 0.10 \pm 0.04$ \\
$C_{\eta' K_S}$ & $-0.16 \pm 0.07 \pm 0.03$ & $0.01 \pm 0.07 \pm 0.05$ \\
\hline \hline
\end{tabular}
\end{center}
\end{table}

The greatest range of $\Delta S_{\eta' K_S}$ is obtained for $C_{\eta' K_S} =
0$.  For the inner ellipse in Fig.\ 1, based on Eq.\ (\ref{boundR3}), one finds
\beq\label{restrictive_bound}
-0.046 < \Delta S_{\eta' K_S} < 0.094~~~,
\eeq
while for the outer ellipse based on Eq.\ (\ref{boundR1}), the limits are
\beq\label{conservative_bound}
-0.133 < \Delta S_{\eta' K_S} < 0.152~~~,
\eeq

Note that the conservative bound \eqref{conservative_bound} uses only SU(3)
symmetry. In obtaining the more restrictive bound \eqref{restrictive_bound}
further dynamical assumptions were made: that the annihilation-like
amplitudes $pa$ and $e$~\cite{GHLR} can be neglected (this can be
justified by taking the $m_b\to \infty$ limit
\cite{Bauer:2004ck,Williamson:2006hb}) and furthermore that the
singlet annihilation-like amplitudes $c_s$ and $s_0$~\cite{Williamson:2006hb}
that depend on the gluonic content of $\eta'$ can be
neglected (the latter do not vanish in the $m_b\to \infty$ limit,
while it is not clear whether or not they are small numerically
\cite{Williamson:2006hb}). The explicit calculations in QCD
Factorization \cite{Beneke:2003zv}, SCET \cite{Williamson:2006hb},
and a model for final state interactions~\cite{Cheng:2005bg}
give results that lie well within both of the above ranges.

\begin{table}
\caption{Time-dependent asymmetries in $B^0 \to \pi^0 K_S$.
\label{tab:pizasyms}}
\begin{center}
\begin{tabular}{c c c} \hline \hline
Parameter & BaBar \cite{Aubert:2006ad} & Belle \cite{Hazumi} \\ \hline
$S_{\pi^0 K_S}$ & $0.33 \pm 0.26 \pm 0.04$ & $0.33 \pm 0.35 \pm 0.08$ \\
$C_{\pi^0 K_S}$ & $0.20 \pm 0.16 \pm 0.03$ & $0.05 \pm 0.14 \pm 0.05$ \\
\hline \hline
\end{tabular}
\end{center}
\end{table}

\section{BOUNDS FOR $S_{\pi^0 K_S}$ and $C_{\pi^0 K_S}$}

We next turn to $B\to \pi^0K_S$ decay.  Measured asymmetries are summarized in
Table \ref{tab:pizasyms}.  The analysis is similar to the one
presented above, with the details given in \cite{Gronau:2003kx}. Using the same
notation as for $B\to \eta' K_S$ we have
\beq\label{SU3}
\sum_f a_f A_f= A(B^0 \to \pi^0\pi^0) + A(B^0 \to K^+ K^-)/\s,
\eeq
so that \eqref{R} gives now
\beq
{\cal R}^2_{\pi/K} \equiv \frac{\bl^2\,[|A_{\pi\pi} +
A_{KK}/\s|^2 + |\bar A_{\pi\pi} +
\bar A_{KK}/\s|^2]}{|A_{\pi K}|^2 + |\bar A_{\pi K}|^2}~.
\eeq
As in \cite{Gronau:2003kx} we now distinguish two cases:
\begin{enumerate}
\item
Neglect the $1/m_b$ suppressed $B^0 \to K^+ K^-$ amplitude for which the
experimental value is $\bb(B^0 \to K^+ K^-) = (0.07 ^{+0.12}_{-0.11})
\cdot 10^{-6}$ \cite{HFAG}.
Then with $\bb(B^0 \to \pi^0 \pi^0) = (1.31 \pm 0.21) \times
10^{-6}$ and $\bb(B^0\to \pi^0 K^0) = (10.0 \pm 0.6)\times 10^{-6}$, we find
\beq
\begin{split}
\label{eqn:noK}
{\cal R}_{\pi/K} = \bl \sqrt{\frac{\bar{\cal B}(B^0 \to \pi^0\pi^0)}
{\bar{\cal B}(B^0 \to \pi^0K^0)}} =(8.3\pm 0.7)\cdot 10^{-2},
\end{split}
\eeq
to be compared with ${\cal R}_{\pi/K} =(9.1 \pm 1.2) \cdot 10^{-2}$ in
\cite{Gronau:2003kx}.
\item
Keeping $A(B^0 \to K^+ K^-)$ increases the error on  ${\cal R}_{\pi/K}$ which
now lies in the range
\beq
\label{eqn:K}
{\cal R}_- \le {\cal R}_{\pi/K} \le {\cal R}_+~,
\eeq
where
\beq
{\cal R}_{\pm} \equiv
\bl\left(\sqrt{\frac{\bar{\cal B}(B^0 \to \pi^0\pi^0)}
{\bar{\cal B}(B^0 \to \pi^0K^0)}} \pm
\sqrt{\frac{\bar{\cal B}(B^0 \to K^+K^-)}
{2\bar{\cal B}(B^0 \to \pi^0K^0)}}\right) \equiv {\cal R}_{\pi/K}
(1 \pm r)~~.
\eeq
With $\bb(B^0 \to K^+ K^-) < (0.224 \times 10^{-6}$ (90\% c.l.) and the
central value of $\bb(B^0 \to \pi^0 \pi^0)$ we find $r < r_{\rm max} = 0.292$.
Then the lower limit on ${\cal R}_{\pi/K}$ becomes ${\cal R}_- = (0.076)(1 -
r_{\rm max}) = 0.054$, while the upper limit becomes ${\cal R}_+ = (0.090)
(1 + r_{\rm max}) = 0.117$.  These are to be compared with ${\cal R}_- =
0.055$, ${\cal R}_+ = 0.126$ obtained in \cite{Gronau:2003kx} using central
values for $\bar{\cal B}(B^0 \to \pi^0K^0)$, $\bar{\cal B}(B^0 \to \pi^0
\pi^0)$ and the upper limit on $\bar{\cal B}(B^0 \to K^+K^-)$.

\end{enumerate}

The results of these two cases are shown in Figs.\ 2.  A small region
of parameter space near the value $(S,C) = (\sin 2 \beta,0)$ is actually
{\it excluded}, as in the case considered in Ref.\ \cite{Gronau:2003kx} when
the small $B^0 \to K^+ K^-$ decay amplitude was ignored.  Here, a small
region is excluded even when $B^0 \to K^+ K^-$ is taken into account.  This is
due in part to the improved upper bounds on this process but also to the
more restricted range assumed for $\gamma$: $52^\circ \le \gamma \le
74^\circ$ \cite{CKMfitter} compared with $38^\circ \le \gamma \le 80^\circ$
taken in Ref.\ \cite{Gronau:2003kx}.

\begin{figure}
\includegraphics[height=3.3in]{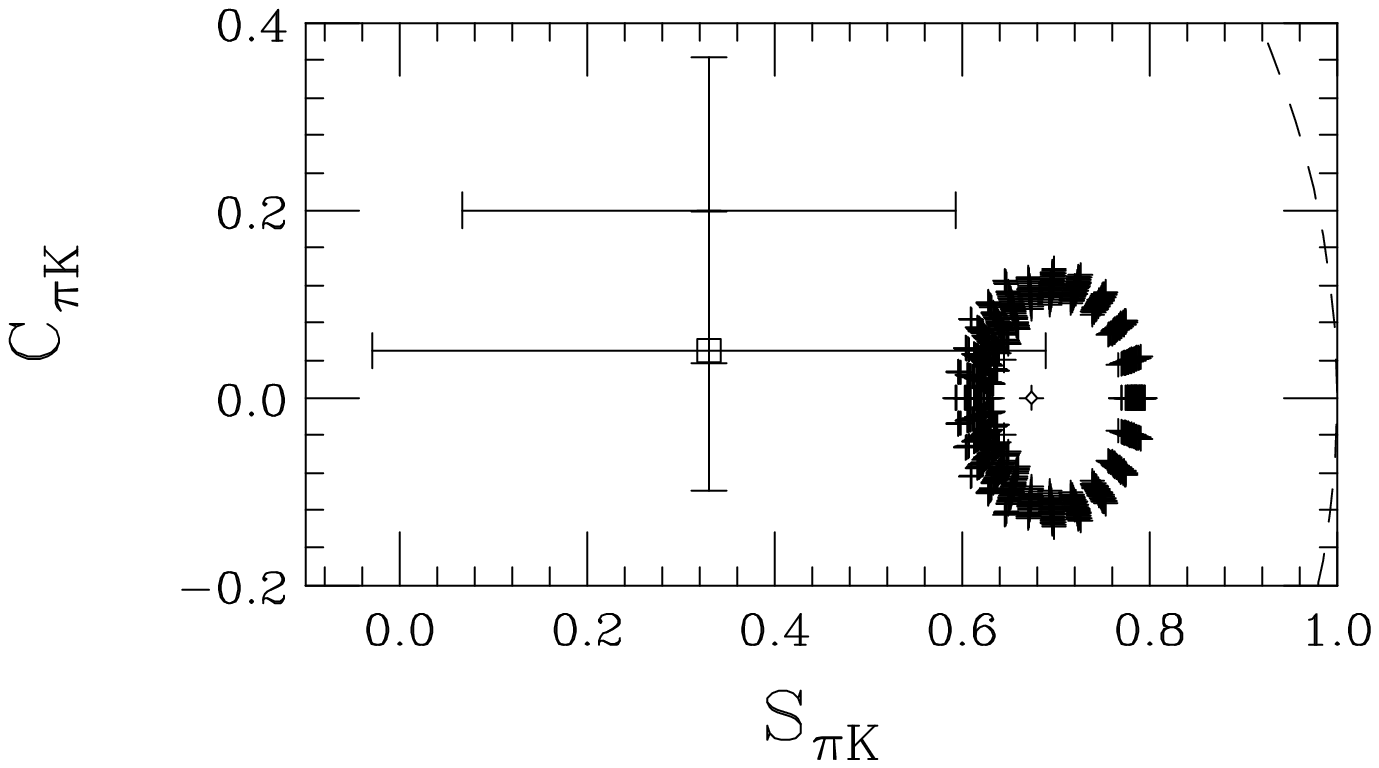}\\[3pt]
\includegraphics[height=3.3in]{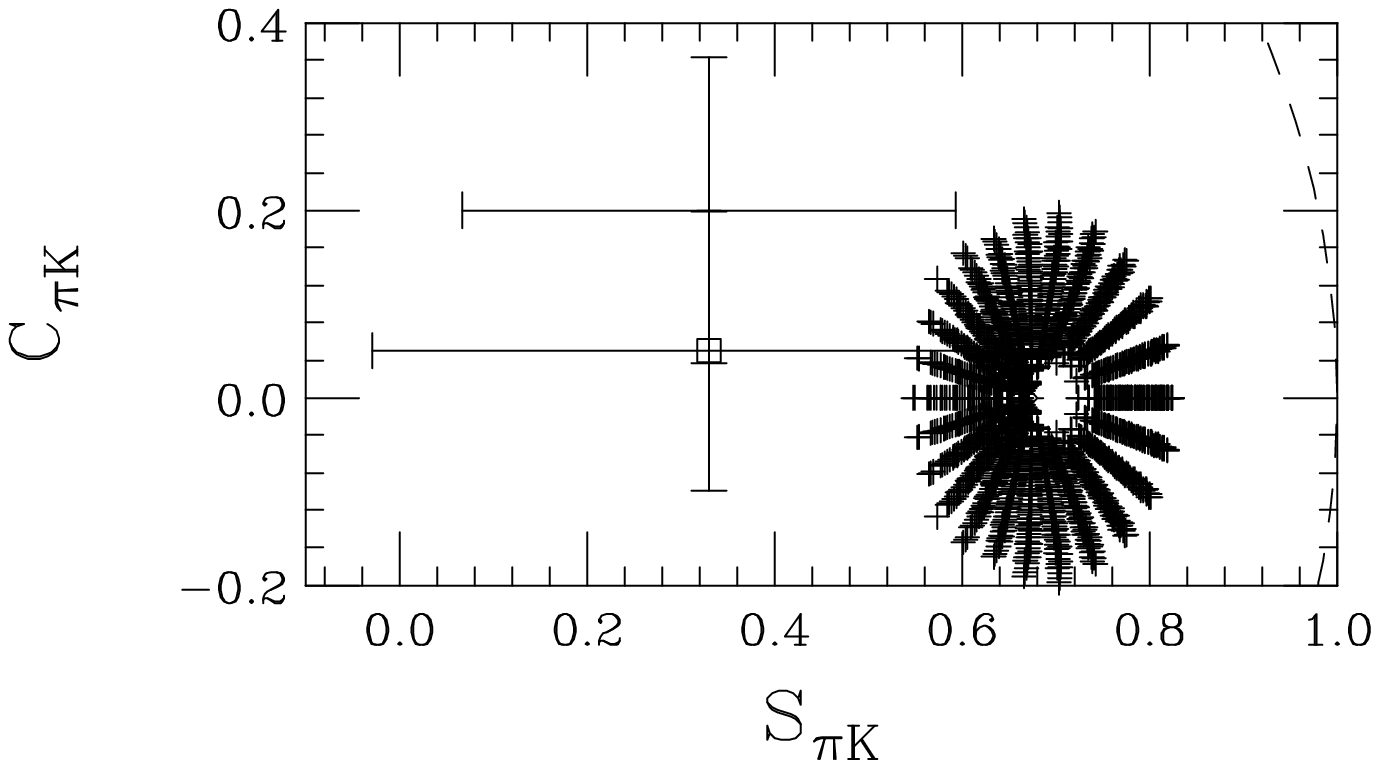}
\caption{Top: points in the $S_{\pi K}$--$|C_{\pi K}|$ plane satisfying $\pm 1
\sigma$ limits (\ref{eqn:noK}) on the ratio $R_{\pi/K}$, with the small $B^0
\to K^+ K^-$ contribution ignored.  The small plotted point denotes the
pure-penguin value $S_{\pi K} = \sin2 \beta$, $C_{\pi K} = 0$.  Points with
experimental errors denote values from BaBar \cite{Aubert:2006ad} (plain point)
and Belle \cite{Hara} (small square).  The dashed arc denotes the boundary of
allowed values:  $S_{\pi K}^2 + C_{\pi K}^2 \le 1$.  Bottom: small $B^0 \to K^+
K^-$ contribution included.
\label{fig:scans}}
\end{figure}

\section{SUMMARY}

SU(3) bounds on the time-dependent CP asymmetries in $B^0 \to \eta' K_S$ and
$B^0 \to \pi^0 K_S$ continue to improve as one incorporates improved bounds on
rare $B^0\to \pi^0\eta^{(')},~\eta^{(')}\eta^{(')}$, and $K^+K^-$ decay
branching ratios.  The bounds presented in this work will thus reach their
minimal values once all the above decay branching ratios are measured.  These
minimal bounds can be estimated using theoretical predictions for $B^0 \to
\pi^0\eta^{(')},~\eta^{(')}\eta^{(')}$, and $K^+K^-$ within QCD Factorization
\cite{Beneke:2003zv}, SCET \cite{Williamson:2006hb} and perturbative QCD (PQCD)
\cite{Wang:2005bk}.  While the central values in these calculations are
typically smaller than the current experimental upper bounds, their theoretical
uncertainties are large, permitting values close to these bounds.  For example,
gluonic contributions to $B\to \eta^{(')}$ form factors may enhance the
relevant branching ratios.  Global SU(3) fits for $B$ decays into two
pseudoscalars obtain values which are within errors near the upper bounds
\cite{CGR,Fu:2003fy,Chiang:2004nm}.  A first indication that the actual
branching ratios are not far below current bounds is the measurement
$\b(B^0\to\pi^0\eta')=(1.5^{+0.7}_{-0.3})\times 10^{-6}$, lying significantly
higher than central values calculated in QCD Factorization and PQCD.  This may
indicate that the bounds \eqref{boundR1} and \eqref{boundR3} will not improve
significantly in the future.

The present constraint on the region around $(S,C) = (\sin 2 \beta,0)$
consistent with the Standard Model is shown in Figs.\ \ref{fig:maxup0608} and
\ref{fig:scans}.  With the new measurements the experimental deviations from
the Standard Model for $B^0 \to \eta' K_S$ have decreased, while those for
$B^0 \to \pi^0 K_S$ are not yet statistically compelling.
  
\section*{Acknowledgments}

We thank Bill Ford, Fernando Palombo, Jim Smith, and Denis Suprun for helpful
comments. M.G. is grateful to the SLAC Theory Group for its kind hospitality.
J.L.R. thanks the Newman Laboratory of Elementary-Particle Physics, Cornell
University, for its kind hospitality during part of this investigation.  This
work was supported in part by the United States Department of Energy, High
Energy Physics Division, through Grant No.\ DE-FG02-90ER40560, under Grants
No.\ DOE-ER-40682-143 and DEAC02-6CH03000, by the Israel Science Foundation
under Grant No.\ 1052/04 and by the German-Israeli Foundation under Grant No.\
I-781-55.14/2003.

\def \ajp#1#2#3{Am.\ J. Phys.\ {\bf#1}, #2 (#3)}
\def \apny#1#2#3{Ann.\ Phys.\ (N.Y.) {\bf#1}, #2 (#3)}
\def \app#1#2#3{Acta Phys.\ Polonica {\bf#1}, #2 (#3)}
\def \arnps#1#2#3{Ann.\ Rev.\ Nucl.\ Part.\ Sci.\ {\bf#1}, #2 (#3)}
\def \art{and references therein}
\def \cmts#1#2#3{Comments on Nucl.\ Part.\ Phys.\ {\bf#1}, #2 (#3)}
\def \cn{Collaboration}
\def \cp89{{\it CP Violation,} edited by C. Jarlskog (World Scientific,
Singapore, 1989)}
\def \econf#1#2#3{Electronic Conference Proceedings {\bf#1}, #2 (#3)}
\def \efi{Enrico Fermi Institute Report No.}
\def \epjc#1#2#3{Eur.\ Phys.\ J.\ C {\bf#1} (#3)  #2}
\def \ib{{\it ibid.}~}
\def \ibj#1#2#3{~{\bf#1}, #2 (#3)}
\def \ijmpa#1#2#3{Int.\ J.\ Mod.\ Phys.\ A {\bf#1}, #2 (#3)}
\def \ite{{\it et al.}}
\def \jhep#1#2#3{JHEP {\bf#1}, #2 (#3)}
\def \jpb#1#2#3{J.\ Phys.\ B {\bf#1}, #2 (#3)}
\def \jpg#1#2#3{J.\ Phys.\ G {\bf#1}, #2 (#3)}
\def \mpla#1#2#3{Mod.\ Phys.\ Lett.\ A {\bf#1} (#3) #2}
\def \nat#1#2#3{Nature {\bf#1}, #2 (#3)}
\def \nc#1#2#3{Nuovo Cim.\ {\bf#1}, #2 (#3)}
\def \nima#1#2#3{Nucl.\ Instr.\ Meth.\ A {\bf#1}, #2 (#3)}
\def \npb#1#2#3{Nucl.\ Phys.\ B~{\bf#1} (#3) #2}
\def \npps#1#2#3{Nucl.\ Phys.\ Proc.\ Suppl.\ {\bf#1}, #2 (#3)}
\def \PDG{Particle Data Group, W.-M. Yao {\it et al.},
\jpg{33}{1}{2006}}
\def \pisma#1#2#3#4{Pis'ma Zh.\ Eksp.\ Teor.\ Fiz.\ {\bf#1},
#2 (#3) [JETP
Lett.\ {\bf#1}, #4 (#3)]}
\def \pl#1#2#3{Phys.\ Lett.\ {\bf#1}, #2 (#3)}
\def \pla#1#2#3{Phys.\ Lett.\ A {\bf#1}, #2 (#3)}
\def \plb#1#2#3{Phys.\ Lett.\ B {\bf#1} (#3) #2}
\def \prl#1#2#3{Phys.\ Rev.\ Lett.\ {\bf#1} (#3) #2}
\def \prd#1#2#3{Phys.\ Rev.\ D\ {\bf#1} (#3) #2}
\def \prp#1#2#3{Phys.\ Rep.\ {\bf#1}, #2 (#3)}
\def \ptp#1#2#3{Prog.\ Theor.\ Phys.\ {\bf#1} (#3) #2}
\def \rmp#1#2#3{Rev.\ Mod.\ Phys.\ {\bf#1}, #2 (#3)}
\def \yaf#1#2#3#4{Yad.\ Fiz.\ {\bf#1}, #2 (#3) [Sov.\
J.\ Nucl.\ Phys.\
{\bf #1}, #4 (#3)]}
\def \zhetf#1#2#3#4#5#6{Zh.\ Eksp.\ Teor.\ Fiz.\ {\bf #1},
#2 (#3) [Sov.\
Phys.\ - JETP {\bf #4}, #5 (#6)]}
\def \zpc#1#2#3{Zeit.\ Phys.\ C {\bf#1} (#3) #2}
\def \zpd#1#2#3{Zeit.\ Phys.\ D {\bf#1}, #2 (#3)}

\end{document}